\documentclass[final,5p,times,twocolumn]{elsarticle}

\usepackage[pscoord]{eso-pic}

\usepackage{graphicx}

\usepackage{amssymb}
\usepackage{amsmath}
\usepackage{gensymb}
\usepackage{enumitem}

\usepackage{lineno}

\biboptions{sort&compress}

\journal{Nuclear Instruments and Methods in Physics Research B }

\begin{document}

\begin{frontmatter}



\title{Total Reaction Cross Sections in CEM and MCNP6 at Intermediate Energies}


\author[label1,label2]{Leslie M. Kerby}   \ead{leslie31415@gmail.com}
\author[label1]{Stepan G. Mashnik}
\address[label1]{Los Alamos National Laboratory, Los Alamos, NM 87545, USA}
\address[label2]{University of Idaho, Moscow, ID 83844, USA}

\begin{abstract}
Accurate total reaction cross section models are important to achieving reliable predictions from spallation and transport codes. The latest version of the Cascade Exciton Model (CEM) as incorporated in the code CEM03.03, and the Monte Carlo N-Particle transport code (MCNP6), both developed at Los Alamos National Laboratory (LANL), each use such cross sections. Having accurate total reaction cross section models in the intermediate energy region ($\sim$50 MeV to $\sim$5 GeV) is very important for different applications, including analysis of space environments, use in medical physics, and accelerator design, to name just a few. The current inverse cross sections used in the preequilibrium and evaporation stages of CEM are based on the Dostrovsky {\it et al.} model, published in 1959. Better cross section models are available now. Implementing better cross section models in CEM and MCNP6 should yield improved predictions for particle spectra and total production cross sections, among other results. Our current 
results indicate this is, in fact, the case.

\end{abstract}


\end{frontmatter}



\section{Introduction}
\label{sec_intro}



Total reaction cross section models have a significant impact on the predictions and accuracy of spallation and transport codes. The latest version of the Cascade Exciton Model (CEM)~\cite{CEM1983,ICTP-IAEA} as incorporated in the code CEM03.03~\cite{ICTP-IAEA,CEM03}, and the Monte Carlo N-Particle transport code (MCNP6)~\cite{MCNP6}, both developed at Los Alamos National Laboratory (LANL), each use such cross sections for different purposes. While total reaction cross sections are used throughout the transport and spallation models, there are two main utilizations. MCNP6 uses total reaction cross sections to determine where a reaction occurs (through the mean-free path length), and then with what nucleus the projectile interacts with, and lastly what type of interaction it is (inelastic or elastic). CEM uses total reaction cross sections as {\it inverse} cross sections to predict what the excited nucleus emits. Phenomenological approximations of total reaction cross sections are also used by CEM03.03 as the default option for normalization of all results in the case of reactions induced by protons and neutrons,
when CEM03.03 is used as a stand alone code, outside any transport codes; 
see details in Refs.~\cite{ICTP-IAEA,CEM03}. 

Having accurate total reaction cross section models in the intermediate energy 
region ($\sim$50 MeV to $\sim$5 GeV) is important for many different applications. 
Applications in space include astronaut radiation dosage, electronics malfunction 
analysis, structural materials analysis, and Galactic Cosmic Rays (GCRs) shielding. 
Medical applications include hadron therapy for cancer \cite{Protons}, 
radiation shielding, medical isotope production,
and high-radiation environment dosimetry. Other applications include accelerator design and simulation. In addition, implementing better inverse cross sections in CEM should provide more reliable predictions; that is, our current work should be useful also from an academic point of view, allowing us to better understand the mechanisms of nuclear reactions. Lastly, the 2008-2010 IAEA Benchmark of Spallation Models recommended an improvement to CEM’s ability to predict the production of energetic light fragments~\cite{SecondAdvancedWorkshop,IAEABenchmark}. Our improvement of the inverse cross sections used by CEM03.03 addresses directly this point, both for a better description of light fragments, but also of nucleons.

The current inverse cross sections used in the preequilibrium and evaporation stages of CEM are based on the Dostrovsky {\it et al.} model, published in 1959 \cite{Dostrovsky}. (For more information about the stages of CEM in its model of spallation reactions, see Ref.~\cite{ICTP-IAEA,NIMA2014, CEM03}.) Better total reaction ({\it inverse}) cross section models are available now \cite{NASAp, NASAn, NASAl, Kox, Townsend, Shen, Takechi, Iida, Ingemarsson, Tsang90, Kalbach, BP}. 

MCNP6 uses an update of the Barashenkov and Polanski (B\&P) cross section model \cite{BP} as described briefly in \cite{BP-last, SARE4REP} to calculate the mean-free path length for neutrons, protons, and light fragments up to $^4$He. It uses a parameterization based on a geometric cross section for light fragments above $^4$He. Implementing better cross section models in CEM and MCNP6 should yield improved results of particle spectra and total production cross sections, among other results. Our current results, upgrading the inverse cross section model in the preequilibrium stage of CEM, prove that this is, in fact, the case.

\begin{figure} [h]
\begin{center}
\includegraphics[width=3.5in]{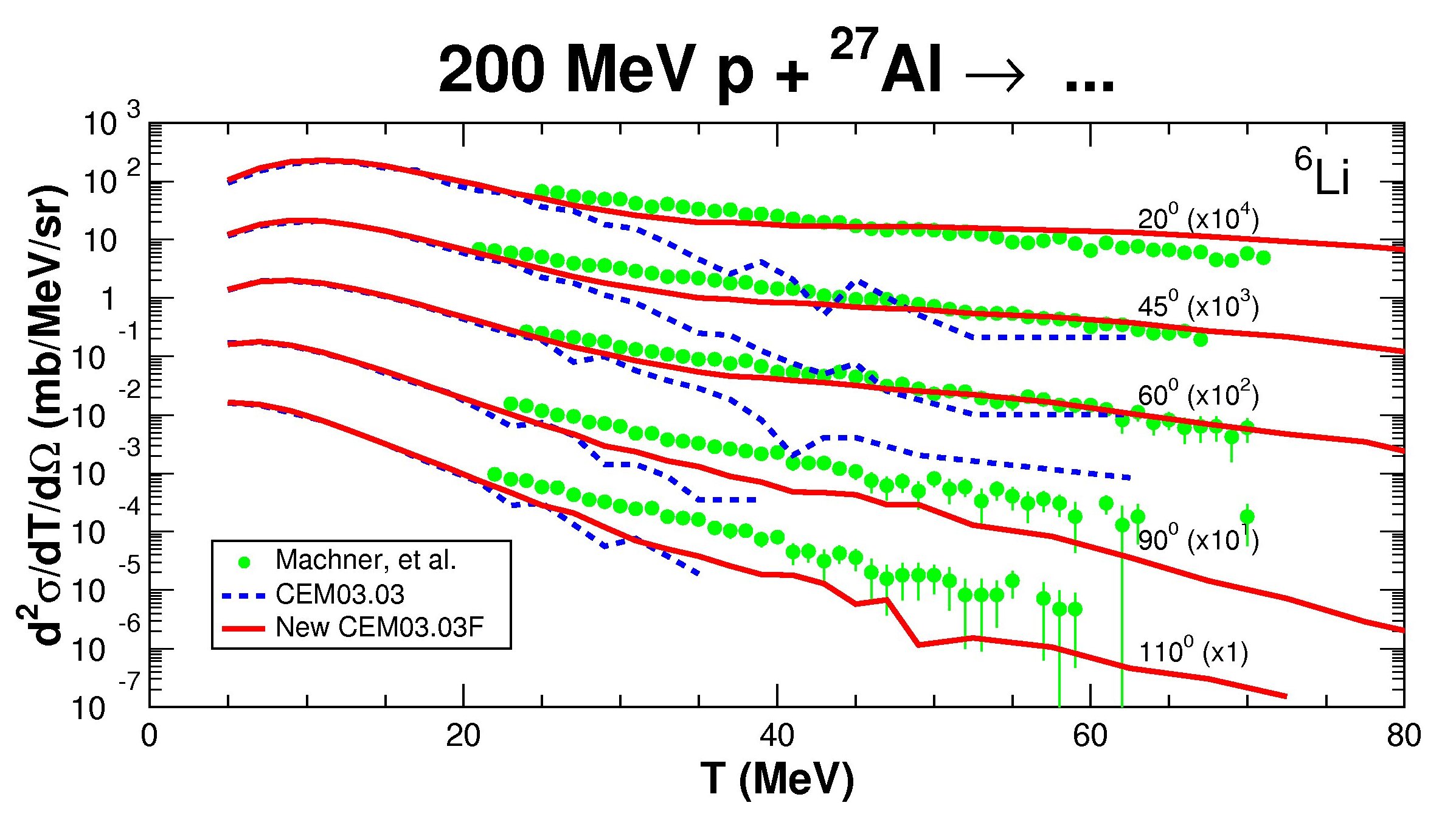}
\caption[]{Comparison of experimental data by Machner {\it et al.} \cite{Machner} (green circles) with results from the unmodified CEM03.03 (blue dotted lines) and the modified--MEM CEM03.03, we refer to here as CEM03.03F \cite{ANS2014, ND2013} (red solid lines) for $^{27}$Al(p,$^6$Li)X with incident proton energy of 200 MeV and for emission angles of 20\degree, 45\degree, 60\degree, 90\degree, and 110\degree.}
\label{fig:p200AlLi6}
\end{center}
\end{figure}

\begin{figure} [h]
\begin{center}
\includegraphics[width=3.5in]{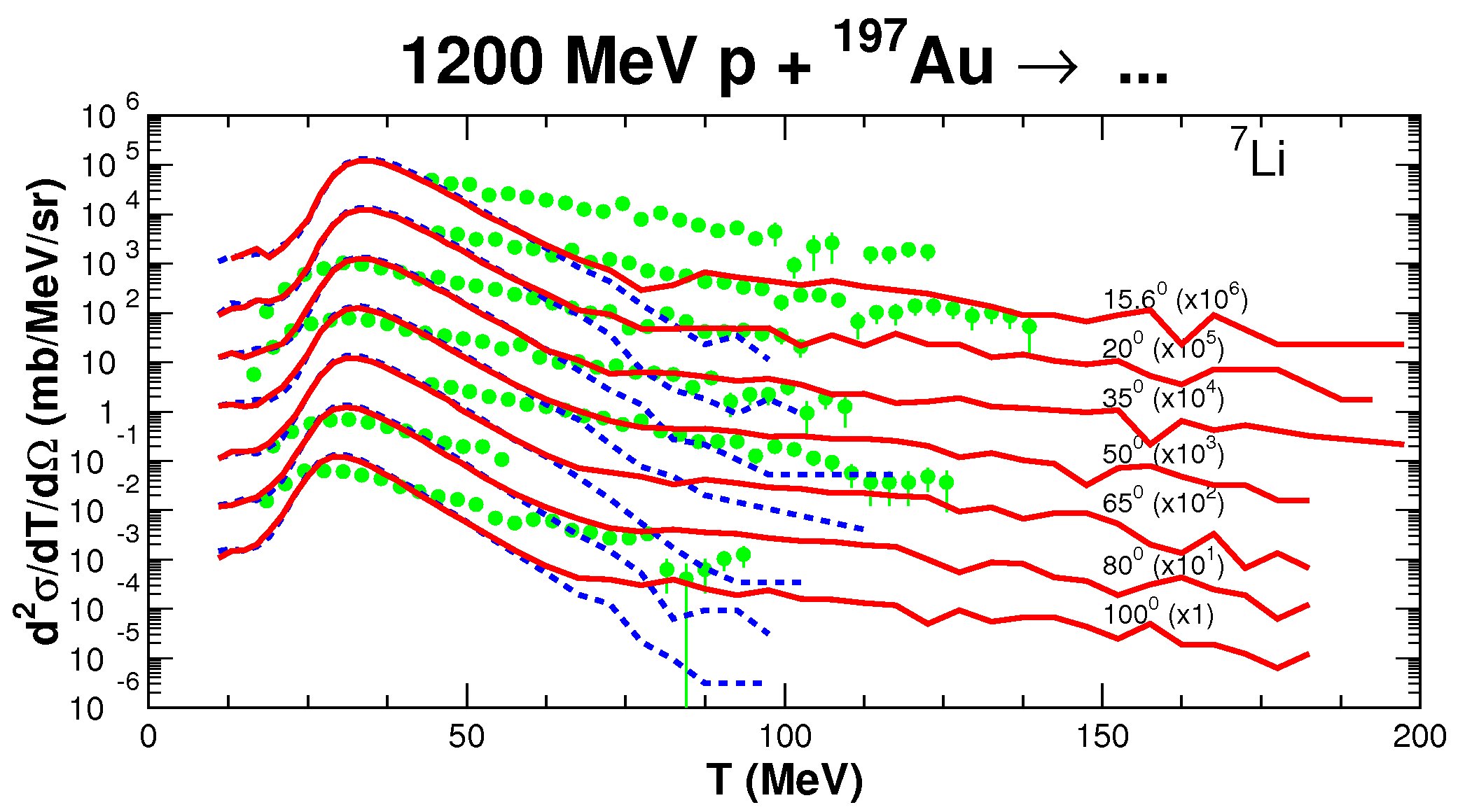}
\caption[]{Comparison of experimental data by Budzanowski {\it et al.} \cite{Budzanowski} (green circles) with results from the unmodified CEM03.03 (blue dashed lines) and the modified-MEM CEM03.03, we refer to here as CEM03.03F \cite{ANS2014, ND2013} (red solid lines) for $^{197}$Au(p,$^7$Li)X with incident proton energy of 1200 MeV and for emission angles of 15.6\degree, 20\degree, 35\degree, 50\degree, 65\degree, 80\degree, and 100\degree. The 100\degree spectrum (the lower set) is shown unscaled, while the 80\degree, 65\degree, etc., down to 15.6\degree spectra are scaled up by successive factors of 10, respectively.}
\label{fig:p1200AuLi7}
\end{center}
\end{figure}

This cross section development work is part of a larger project aimed at enabling CEM to produce high-energy light fragments \cite{NIMA2014, ANS2014, ND2013}. Figs.~\ref{fig:p200AlLi6} and~\ref{fig:p1200AuLi7} illustrate two examples of results of that project: comparing results from CEM03.03 with an upgraded Modified Exciton Model (MEM) to results from CEM03.03 unmodified. For some reactions we obtained good results (see, e.g., Fig.~\ref{fig:p200AlLi6}), and for other reactions, while our results showed improvement, they could still be better (see, e.g., Fig.~\ref{fig:p1200AuLi7}). We decided to upgrade the inverse cross section models used by CEM, in the preequilibrium stage, to improve such results further. As CEM is the default event generator in MCNP6 in the intermediate energy range, once these results our implemented into MCNP6 (to be completed soon), we should see a corresponding improvement in MCNP6 as well. 

\section{Background}
As mentioned above, the current inverse cross sections in CEM are based on the Dostrovsky {\it et al.} model \cite{Dostrovsky}. It is based on the strong absorption model and its general form is as shown in Eq.~(\ref{eq:Dost}).

\begin{equation}
\sigma_{Dost.} = \pi r_0^2 A^{2/3} \alpha_j(1 - \frac{V_j}{T}) .
\label{eq:Dost}
\end{equation}

The Dostrovsky {\it et al.} model was not intended for use above about 50 MeV/nucleon, and is not very suitable for emission of fragments heavier than $^4$He. Better total reaction cross section models are available today, most notably the NASA model \cite{NASAp, NASAn, NASAl}. The NASA (or Tripathi {\em et al.}) model is also based on the strong absorption model and its general form is shown in Eq.~(\ref{eq:NASA}). $\delta_T$, $X_m$, and $B_T$ are discussed more fully later, and defined in Eq.~(\ref{eq:NASA_details}). The NASA cross section attempts to simulate several quantum-mechanical effects, such as the optical potential for neutrons (with $X_m$) and collective effects like Pauli blocking (through $\delta_T$). (For more details, see Refs.~\cite{NASAp, NASAn, NASAl}.)

\begin{equation}
\sigma_{NASA} = \pi r^2_0 (A_P^{1/3} + A_T^{1/3} + \delta_T )^2 
(1 - R_c \frac{B_T}{T_{cm}})X_m \mbox{ ,}
\label{eq:NASA}
\end{equation}
where
\begin{description}[noitemsep]
	\item[] $r_0$ is a constant related to the radius of a nucleus;
	\item[] $A_P$ is the mass number of the projectile nucleus; 
	\item[] $A_T$ is the mass number of the target nucleus;
	\item[] $\delta_T$ is an energy-dependent parameter;
	\item[] $R_c$ is  a system-dependent Coulomb multiplier;
	\item[] $B_T$ is the energy-dependent Coulomb barrier;
	\item[] $T_{cm}$ is the colliding system center-of-momentum energy;
	\item[] $X_m$ is an optical model multiplier used for neutron-induced reactions.
\end{description}

There are other proposed total reaction cross section models, such as those by Shen, {\em et al.} \cite{Shen}, and Takechi, {\em et al.} \cite{Takechi}, amongst others \cite{BP, Iida, Ingemarsson, Tsang90, Kalbach, Kox, Townsend}. It should be noted that both the Shen model and the Kox model have projectile-target assymetry, as discussed in Ref.~\cite{Sihver2014}. In Ref.~\cite{SihverPHITS}, Sihver {\em et al}. explores a new total reaction cross section used in PHITS: the hybrid Kurotama model. This model is a combination of the Black Sphere model \cite{Iida} and the NASA model \cite{NASAp, NASAn, NASAl}. Ref.~\cite{SihverFLUKA} compares a number of different total reaction cross section models, most notably those in FLUKA, NASA, and several other recently developed models. 

PHITS uses the NASA model as its default total reaction cross section model, but Shen can be specified as an option \cite{SihverFLUKA}. FLUKA uses a modified version of the NASA model as its total reaction cross section model \cite{Anderson}. GEANT4 has the option to use NASA, or a number of other total reaction cross section models such as Shen \cite{Shen} or Sihver \cite{Sihver}, or the Axen-Wellisch \cite{Axen} total reaction cross section parameterizations for high-energy hadronic interactions. See Ref.~\cite{SihverGEANT, GEANTManual} for more details on the total reaction cross section models used in PHITS, FLUKA and GEANT4.

\begin{figure*}[]
\centering
\includegraphics[trim = 1.0in 1.7in 1.0in 1.7in, width=6.0in]{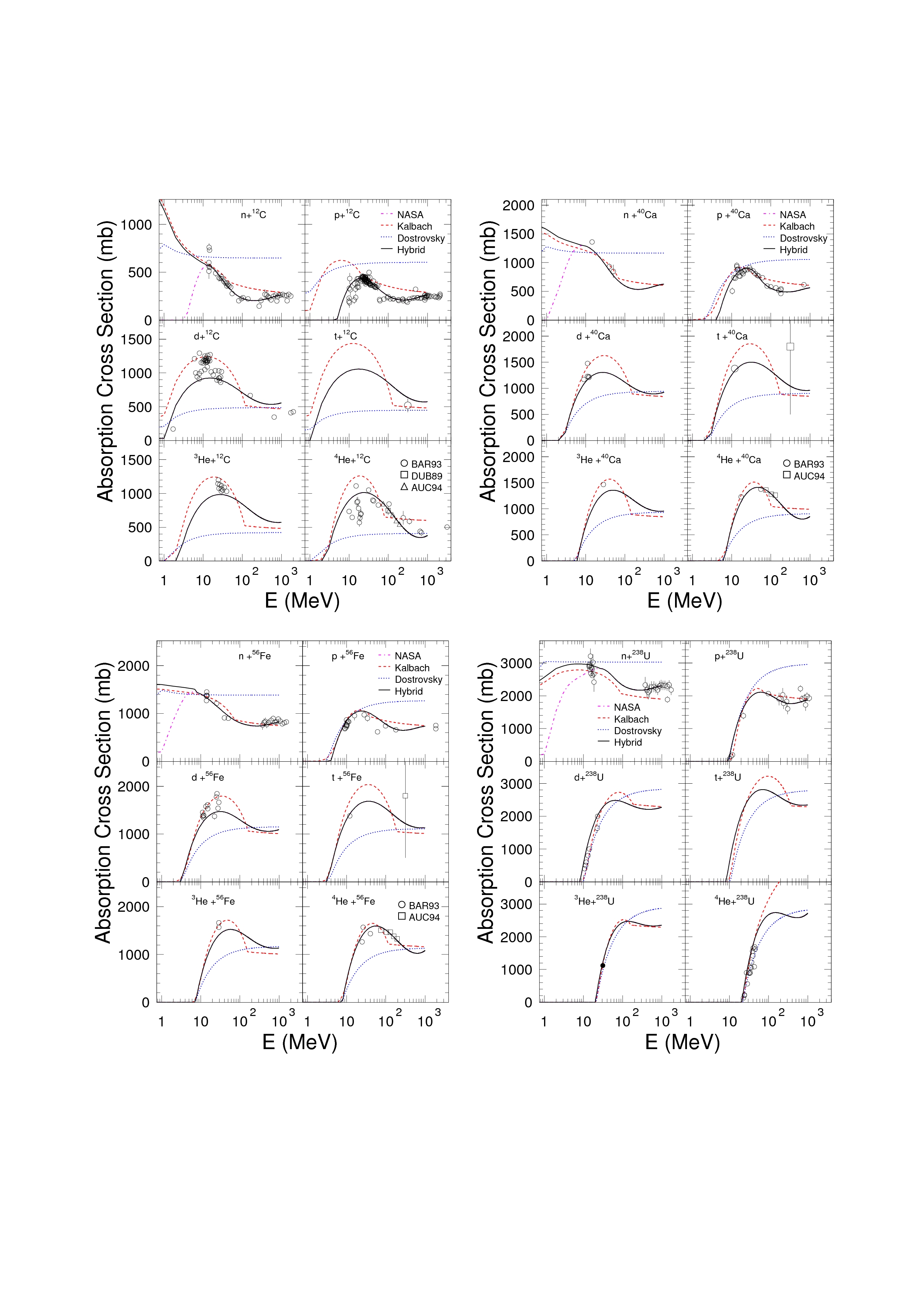}
\caption[]{Absorption (inverse) cross section by energy for various reactions, as calculated in Ref. \cite{SantaFe2002} by the NASA \cite{NASAp, NASAn, NASAl}, Kalbach \cite{Kalbach}, and Dostrovsky {\it et al.} \cite{Dostrovsky} systematics, as well as with a ``Hybrid approach'' suggested in \cite{SantaFe2002} to account for both NASA \cite{NASAn} and Kalbach \cite{Kalbach} systematics, in the case of neutron-induced reactions. ``BAR93'' shows experimental data from Ref.~\cite{BarashenkovTables}; ``DUB89'' shows data from Ref.~\cite{DUB89}; and ``AUC94'' shows data from Ref.~\cite{AUC94}.}
\label{fig:Mashnik}
\end{figure*}

Stepan Mashnik with collaborators \cite{SantaFe2002, ResNote2006} and Dick Prael with coauthors \cite{SARE4REP, SARE4REP-suppl} previously conducted at LANL an extensive comparison of the NASA \cite{NASAp, NASAn, NASAl}, Tsang {\it et al.} \cite{Tsang90}, Dostrovsky {\it et al.} \cite{Dostrovsky}, Barashenkov and Polanski (using their code called CROSEC) \cite{BP}, and Kalbach \cite{Kalbach} systematics for total reaction ({\it inverse}) cross sections. Fig.~\ref{fig:Mashnik} illustrates some results from the study \cite{SantaFe2002}. Their results found that the NASA total reaction cross section model was superior, in general, to the other available models. See Ref.~\cite{SARE4REP,SantaFe2002,ResNote2006,FY14} for details of their findings.

\section{Comparison of Total Reaction Cross Section  Models}
We built in CEM03.03F the NASA model \cite{NASAp, NASAn, NASAl} and the models used in the preequilibrium (labeled as ``Dostrovsky'' in our figures below) and the evaporation (described with the code GEM2 by Furihata~\cite{GEM2thesis}, therefore labeled in our figures below as ``GEM2'') stages of CEM03.03, and also compared reactions to calculations from the Barashenkov and Polanski (B\&P) systematics \cite{BP}, and, for comparison, to two neutron- and proton-induced reaction cross sections calculations by MCNP6 \cite{MCNP6}. Note that MCNP6 uses currently an updated and improved version of the initial Barashenkov and Polanski (B\&P) systematics \cite{BP}, as outlined briefly in Refs. \cite{BP, BP-last}, to simulate the mean-free path length of nucleons in matter.

\subsection{Neutron-Induced Reactions}
\begin{figure}[h]
\centering
\includegraphics[trim = 1.25in 2.25in 1.25in 2.25in, width=3.5in]{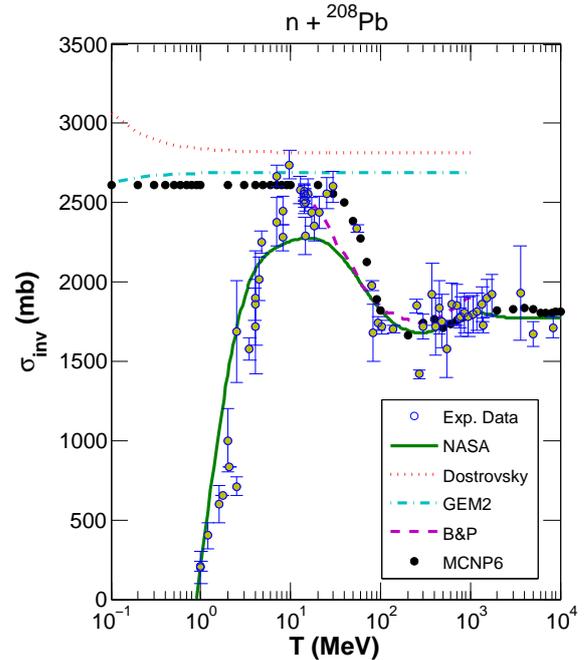}
\caption[]{Reaction cross section for n + $^{208}$Pb, as calculated by the NASA, Dostrovsky {\it et al.}, GEM2, and B\&P models. The black dots are cross section calculations of MCNP6, and the yellow circles are experimental data \cite{Bonner, Beyster1956, Taylor, Pasechnik, Beyster1955, Strizhak, Beghian, Morrison, Degtjarev, Walt, Poze}.}
\label{fig:n+Pb}
\end{figure}

Fig.~\ref{fig:n+Pb} displays the total reaction cross section for n + $^{208}$Pb, as calculated by the NASA, Dostrovsky {\it et al.}, GEM2, and B\&P models, and compared to calculations by MCNP6 and experimental data. There are several things to notice: 1) the Dostrovsky and GEM2 (also a Dostrovsky-based model) both approach asymptotic values very quickly--thus they are not as useful at their constant values, and 2) the NASA model, while much 
better at predicting the total reaction cross section throughout the energy 
region of projectiles, falls to zero at low energies 
in the case of neutrons, where we do not have Coulomb barriers. 
For this reason, we can not use the NASA model as an approximation
for inverse cross sections in the case of low-energy neutrons: 
neutrons are emitted with low energies, too. In the case of low energy
neutrons, we use the Kalbach systematics [20], which proved to be a very
good approximation for the inverse cross section of low-energy
neutrons, as discussed in Ref. [36] and in sub-section 4.1 below.
Note that this problem of neutron cross sections was addressed first for the code CEM2k in Ref.~\cite{SantaFe2002}, by combining the NASA systematics by Tripathi, Cucinota, and Wilson~\cite{NASAp,NASAn,NASAl} and the Kalbach parameterization~\cite{Kalbach} into a FORTRAN routine called {\sf hybrid}. We address this problem here, for our current CEM03.03F code, in a very similar way (see Ref.~\cite{FY14}).

See Ref.~\cite{FY14} for results of other neutron-induced reactions.

\subsection{Proton-Induced Reactions}
\begin{figure}[h]
\centering
\includegraphics[trim = 1.25in 2.5in 1.25in 2.5in, width=3.5in]{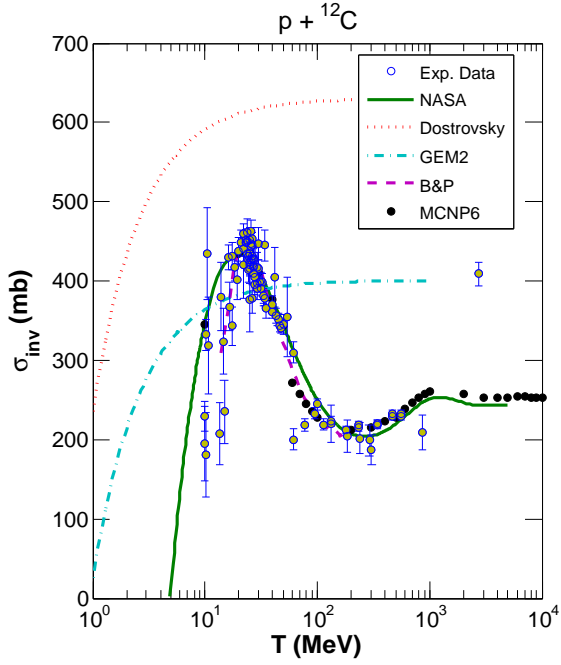}
\caption[]{Reaction cross section for p + $^{12}$C, as calculated by the NASA, Dostrovsky {\it et al.}, GEM2, and B\&P models. The black dots are cross section calculations of MCNP6, and the yellow circles are experimental data \cite{Carlson}.}
\label{fig:p+C}
\end{figure}

Fig.~\ref{fig:p+C} illustrates calculated total reaction cross sections by 
the NASA, Dostrovsky {\it et al.}, GEM2, and B\&P models, compared to calculations 
by MCNP6 and experimental data. The NASA model appears to be superior to the 
Dostrovsky-based models. 
 
As we can see from Figs. 5 and 4 on nucleons, as well as from
examples on complex-particles and fragments heavier than $^4$He
presented below in Figs. 6 and 7, and in numerous figures 
published in Refs. [23, 36-38], the Barashenkov and Polanski approximations also
agree very well with available data. For this reason, the B\&P parametrization was chosen
to be used 
for the calculation of the total reaction cross sections in the transport
code MCNP6 [4], and in several other transport codes, too, as far as we know. 
Howerver, our numerous
current comparisions for various reactions, as well as the voluminous
results published in
Refs. [23, 36-38], show that, on the whole, the NASA approximation agree a litlle
better with most of the available experimental data than the B\&P systematics does.
 
See Ref.~\cite{FY14} for results of other proton-induced reactions.

\subsection{Heavy-Ion Induced Reactions}
We never tested before how CEM03.03 calculates inverse cross sections for light fragments (LF) heavier than $^4$He. We address this question below.

\begin{figure}[h!]
\centering
\includegraphics[trim = 1.25in 1.5in 1.25in 2.9in, width=3.5in]{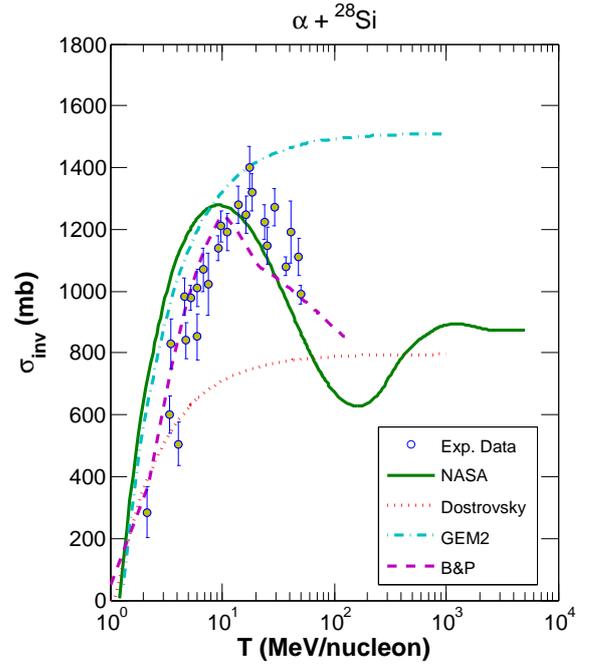}
\includegraphics[trim = 1.25in 2.5in 1.25in 2.9in, width=3.5in]{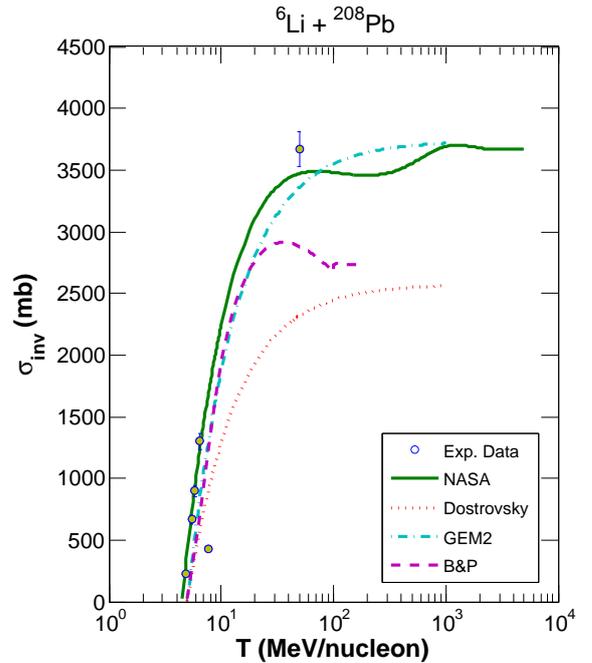}
\caption[]{Reaction cross section for $\alpha$ + $^{28}$Si and $^6$Li + $^{208}$Pb, as calculated by the NASA, Dostrovsky {\it et al.}, GEM2, and B\&P models. The yellow circles are experimental data \cite{Warner, Ugryumov2005, Ugryumov2004, Ingemarsson2000, Baktybaev}.}
\label{fig:4He+28Si}
\end{figure}

Fig.~\ref{fig:4He+28Si} illustrates calculated total reaction cross sections by the NASA, Dostrovsky {\it et al.}, GEM2, and B\&P models for the reactions $\alpha$ + $^{28}$Si and $^6$Li + $^{208}$Pb, compared to experimental data. 

\begin{figure}[h]
\centering
\includegraphics[trim = 1.3in 2.4in 0.7in 2.7in, width=3.8in]{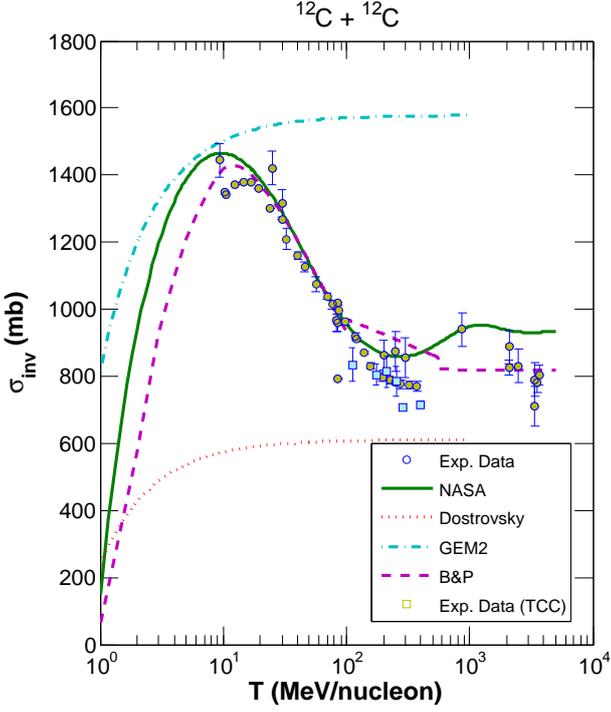}
\caption[]{Reaction cross section for $^{12}$C + $^{12}$C, as calculated by the NASA, Dostrovsky {\it et al.}, GEM2, and B\&P models. The yellow circles are experimental data \cite{BarashenkovTables, Mazarakis, Takechi} and the blue squares are total charge-changing cross section (TCC) measurements \cite{Zeitlin, Golovchenko}.}
\label{fig:12C+12C}
\end{figure}

Fig.~\ref{fig:12C+12C} displays the total reaction cross section for $^{12}$C + $^{12}$C, as calculated by the NASA, Dostrovsky {\it et al.}, GEM2, and B\&P models and compared to experimental data and to measured total charge-changing (TCC) cross sections. TCC cross sections should be $5\%-10\%$ less than total reaction cross sections, as TCC cross sections do not include the neutron removal cross section.

See Ref.~\cite{FY14} for results of other heavy-ion-induced reactions.

We determined that the NASA cross section model fits the experimentally measured data, in general, better than the other models tested.

\section{Implementation of NASA Cross Section Model into CEM03.03F}
The implementation of the NASA cross section model into CEM involved adding Kalbach systematics for low-energy neutrons, updating the emission width calculation, and upgrading the emitted fragment kinetic energy simulation.

\subsection{Kalbach Systematics}
\begin{figure}[h]
\centering
\includegraphics[trim = 1.25in 2.5in 1.25in 2.5in, width=3.5in]{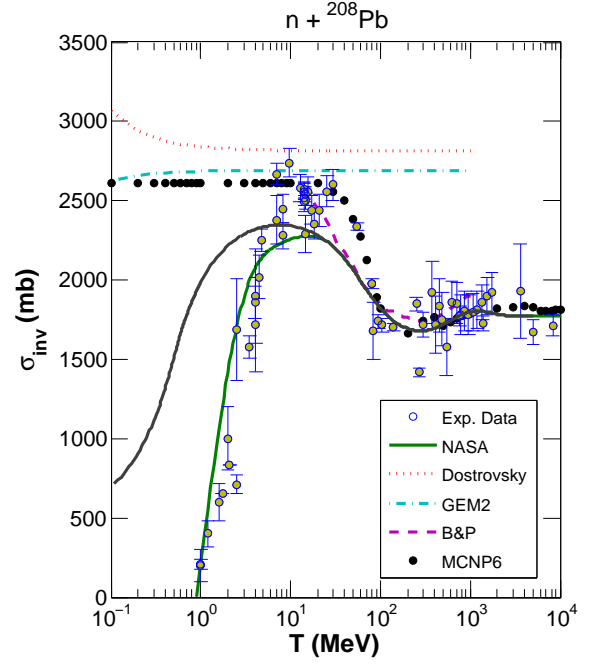}
\caption[]{Reaction cross section for n + $^{208}$Pb, as calculated by the NASA, NASA-Kalbach hybrid (black line), Dostrovsky {\it et al.}, GEM2, and B\&P models, as indicated. The black dots are cross section calculations by MCNP6, and the yellow circles are experimental data from Refs.~\cite{Bonner, Beyster1956, Taylor, Pasechnik, Beyster1955, Strizhak, Beghian, Morrison, Degtjarev, Walt, Poze}.}
\label{fig:n+Pb_Kal}
\end{figure}

We added in CEM03.03F the Kalbach systematics~\cite{Kalbach} to replace the NASA inverse cross sections~\cite{NASAp,NASAn,NASAl} for low-energy neutrons, similar to what was suggested and done in Ref.~\cite{SantaFe2002} for the code CEM2k. Fig.~\ref{fig:n+Pb_Kal} displays the Kalbach systematics implementation for the cross section n + $^{208}$Pb. At around 24 MeV and below, the calculation switches to Kalbach systematics, and uses the NASA model throughout the rest of the neutron-energy range. The Kalbach systematics is scaled to match the NASA model results at the switchpoint so as not to have a large jump. 

As part of the Kalbach systematics implementation in CEM03.03F, switchpoints and scaling factors must be obtained for all possible residual nuclei, by mass number. Ref.~\cite{FY14} provides tables of these.

\subsection{Emission Width, $\Gamma_j$, Calculation}
CEM uses the inverse cross section, $\sigma_j^{inv}$, in determining what particles and/or fragments are emitted from the excited nucleus. We use the total reaction cross section as the best approximation for this inverse cross section. The emission width $\Gamma_j$, or the probability of emitting fragment type $j$, is calculated according to Eq.~(\ref{eq:Gamma_j}). It is dependent upon $\sigma_j^{inv}$ (see more details in Refs.~\cite{CEM1983,ICTP-IAEA,CEM03}). 

\begin{equation}
\begin{split}
\Gamma_{j}(p,h,E) = & \int_{V_j^c}^{E-B_j} 
\frac{2s_j + 1}{\pi^2\hbar^3} \mu_j \Re (p,h) \\
& \times \frac{\omega (p-1,h,E-B_j-T)}{\omega (p,h,E)} T \sigma_j^{inv} (T) dT \mbox{ ,}
\end{split}
\label{eq:Gamma_j}
\end{equation}
where
\begin{description}[noitemsep]
	\item[] $s_j$ is the spin of the emitted particle $j$; 
	\item[] $\mu_j$ is the reduced mass of the emitted particle $j$; 
	\item[] $\omega$ is the level density of the $n$-exciton state; 
	\item[] $B_j$ is the binding energy; 
	\item[] $V_j^c$ is the Coulomb barrier; 
	\item[] $T$ is the kinetic energy of the emitted particle $j$;
	\item[] $\sigma^{inv}_j$ is the inverse cross section; and
	\item[] $\Re$ creates zero probability of emission if the number of 
particle excitons is less than the number of nucleons of particle~$j$. 
\end{description}
Eq.~(\ref{eq:Gamma_j}) is written in its simplest form, as is valid for neutrons and protons only. An extension of Eq.~(\ref{eq:Gamma_j}) for the case of complex particles and light fragments (LF) is described in detail in Ref.~\cite{CEM1983}.

In the ``standard'' (i.e., ``old,'' for this study) calculation by CEM03.03, performed with a FORTRAN routine called {\sf gamagu2}, therefore referred to below as ``gamagu2,'' the Dostrovsky {\it et al.} form of the inverse cross section is simple enough so that for neutrons and protons this integral can be done analytically. However, for complex particles, 
the level density, $\omega$, 
becomes too complicated (see details in Refs.~\cite{CEM1983,ICTP-IAEA,CEM03}), therefore the integral is evaluated numerically. In this case, a 6-point Gaussian quadrature is used when the exciton number is 15 or less, and a 6-point Gauss-Laguerre quadrature is used when the number of excitons is over 15. We will soon see why the two different integration methods are needed.

In our current calculations we adopt here for CEM03.03F (performed with a FORTRAN routine called {\sf gamagu3}, hereafter referred to as ``gamagu3''), the NASA form of the cross section is too complicated and the integral is always calculated numerically. We use an 8-point Gaussian quadrature when the number of excitons is 15 or less, and an 8-point Gauss-Laguerre quadrature when the number of excitons is greater than 15.

The partial transmission probability $\lambda_j$, or the probability that a particle or a fragment of the type $j$ will be emitted with kinetic energy $T$, is equal to the integrand of Eq.~(\ref{eq:Gamma_j}). For the emission of LF this is equal to
\begin{equation}
\begin{split}
\lambda_{j}(p,h,E,T) = & \gamma_j \frac{2s_j + 1}{\pi^2\hbar^3} \mu_j \Re (p,h) \\
	& \times \frac{\omega (p-p_j,h,E-B_j-T)}{\omega (p,h,E)} \\
	& \times \frac{\omega (p_j,0,T+B_j)}{g_j} T \sigma_j^{inv} (T) \mbox{ ,}
\end{split}
\label{eq:lambda_j}
\end{equation}

where

\begin{equation}
g_j = \frac{V (2\mu_j)^{3/2}}{4\pi^2\hbar^3} (2s_j + 1) (T+B_j)^{1/2} \mbox{ .}
\label{eq:g_j}
\end{equation}
See Ref.~\cite{Wu} for details on Eq.~(\ref{eq:g_j}). For completeness sake, we write Eq.~(\ref{eq:lambda_j}) (and also Eq.~(\ref{eq:lambda_j_NASA}) below in the text) in their ``complete form,'' as they should look in the case of complex particles and LF, but not in their ``simplest'' version needed only for nucleons as exemplified by Eq.~(\ref{eq:Gamma_j}).
 
\begin{figure}[ht!]
\centering
\includegraphics[trim = 1.25in 2.in 1.25in 3.25in, width=3.5in]{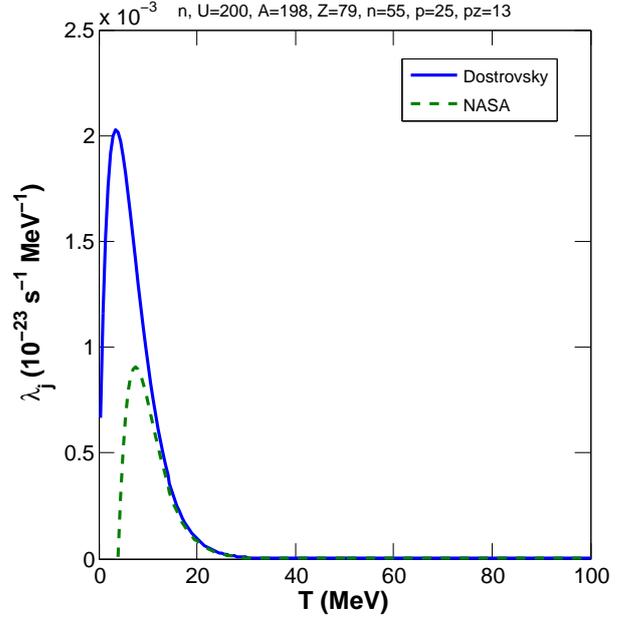}
\includegraphics[trim = 1.25in 2.75in 1.25in 3.in, width=3.5in]{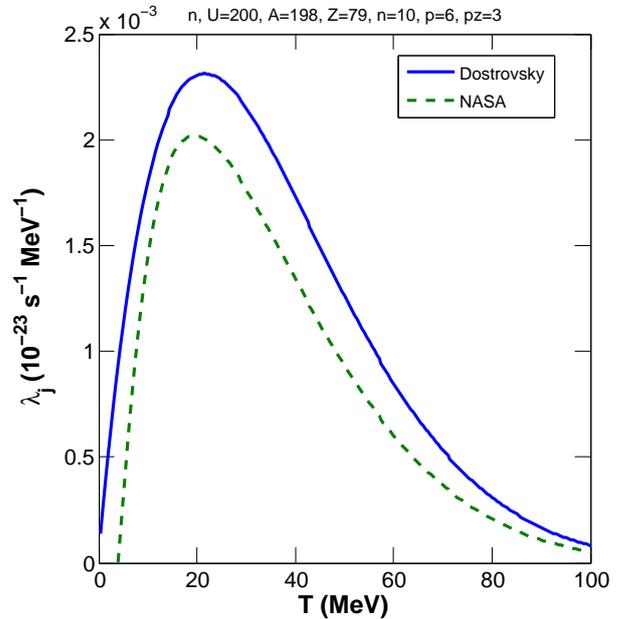}
\caption[]{$\lambda_j$ as a function of the kinetic energy of the emitted neutron, from an excited $^{198}$Au nucleus with $U$ = 200 MeV and 55 excitons, 25 particle excitons, and 13 charged particle excitons (top plot) and 10 excitons, 6 particle excitons, and 3 charged particle excitons (bottom plot).}
\label{fig:lambda_j}
\end{figure}

As an example, Fig.~\ref{fig:lambda_j} shows $\lambda_j$ for the emission of neutrons from a $^{198}$Au excited nucleus, with an internal nucleus energy $U$ of 200 MeV, using either the Dostrovsky {\it et al.} or NASA cross section. The top plot is for 55 excitons and the bottom plot is for 10 excitons. Notice that for high exciton number, $\lambda_j$ becomes more concentrated in the low-energy region. Table~\ref{tab:quad} displays the abscissas for an 8-point Gaussian and an 8-point Gauss-Laguerre quadrature. For a small number of excitons ($\leq 15$) the Gaussian quadrature performs adequately. However, we see that in the 55-exciton case the $\lambda_j$ becomes negligible by about 30 MeV, requiring a different integration method. For high-exciton number the Gauss-Laguerre integration method is a much better choice than the simple Gaussian. 

\begin{table}[h]
\centering
\caption{8-point Gaussian and Gauss-Laguerre sampling points}
\begin{tabular}{cc}
\hline\hline 
 8-pt Gaussian & 8-pt Gauss-Laguerre \\
\hline
3.84 MeV	&	0.428 MeV \\
19.7 MeV	&	2.27 MeV \\
45.9 MeV	&	5.66 MeV \\
79.0 MeV	&	10.7 MeV \\
114. MeV	&	17.7 MeV \\
148. MeV	& 	27.1 MeV \\
174. MeV	&	39.6 MeV \\
190. MeV	&	57.5 MeV \\
\hline\hline 
\end{tabular}
\label{tab:quad}
\end{table}

\begin{figure}[h]
\centering
\includegraphics[trim = 1.25in 2.5in 1.25in 2.5in, width=3.5in]{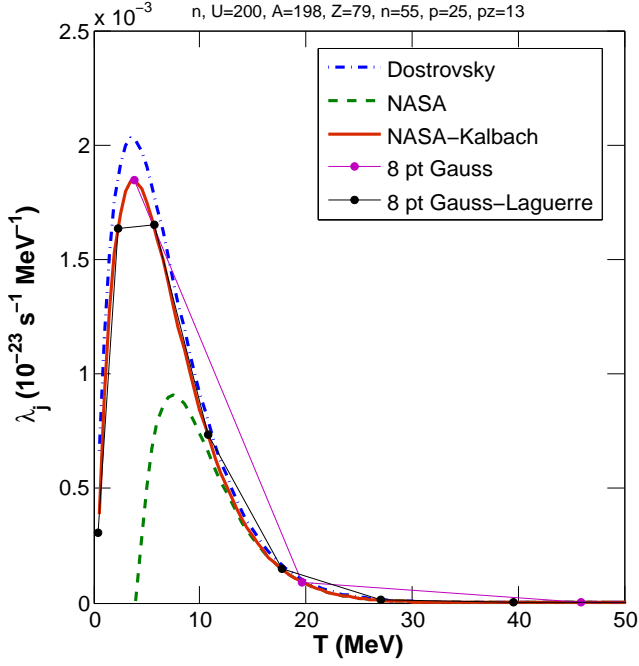}
\caption[]{$\lambda_j$ as a function of the kinetic energy of the emitted neutron, from an excited $^{198}$Au nucleus with $U$ = 200 MeV and 55 excitons, 25 particle excitons, and 13 charged particle excitons.}
\label{fig:lambda_j_Kal}
\end{figure}

Fig.~\ref{fig:lambda_j_Kal} shows a comparison of the simple Gaussian and Gauss-Laguerre quadratures for 55 excitons. This figure also displays $\lambda_j$ for the NASA-Kalbach cross section. Notice that the NASA-Kalbach has much higher values of $\lambda_n$ at the low end of the spectrum than the pure NASA. The purple dots are the 8-pt Gaussian quadrature and the black dots are the 8-pt Gauss-Laguerre quadrature. The Gaussian was exceptionally fortunate in that it struck the peak with its one low-end point. However, this leads to significant overestimation of $\lambda_j$ down the tail. The Gauss-Laguerre underestimates the peak but then overestimates slightly along the tail. Even though it is clear this is not a very close fitting of $\lambda_j$, changing to a 10-pt Gauss-Laguerre only yielded a 0.2\% difference. A future project could include investigating the behavior of $\lambda_j$ across the variable landscape, and implementing an adaptive quadrature scheme. However, whatever numerical integration method we use, it must be fast as this integral is calculated hundreds of times for every event, and therefore billions of times for a typical simulation.

\begin{figure}[]
\centering
\includegraphics[trim = 1.75in 3.in 1.75in 3.5in, width=2.5in]{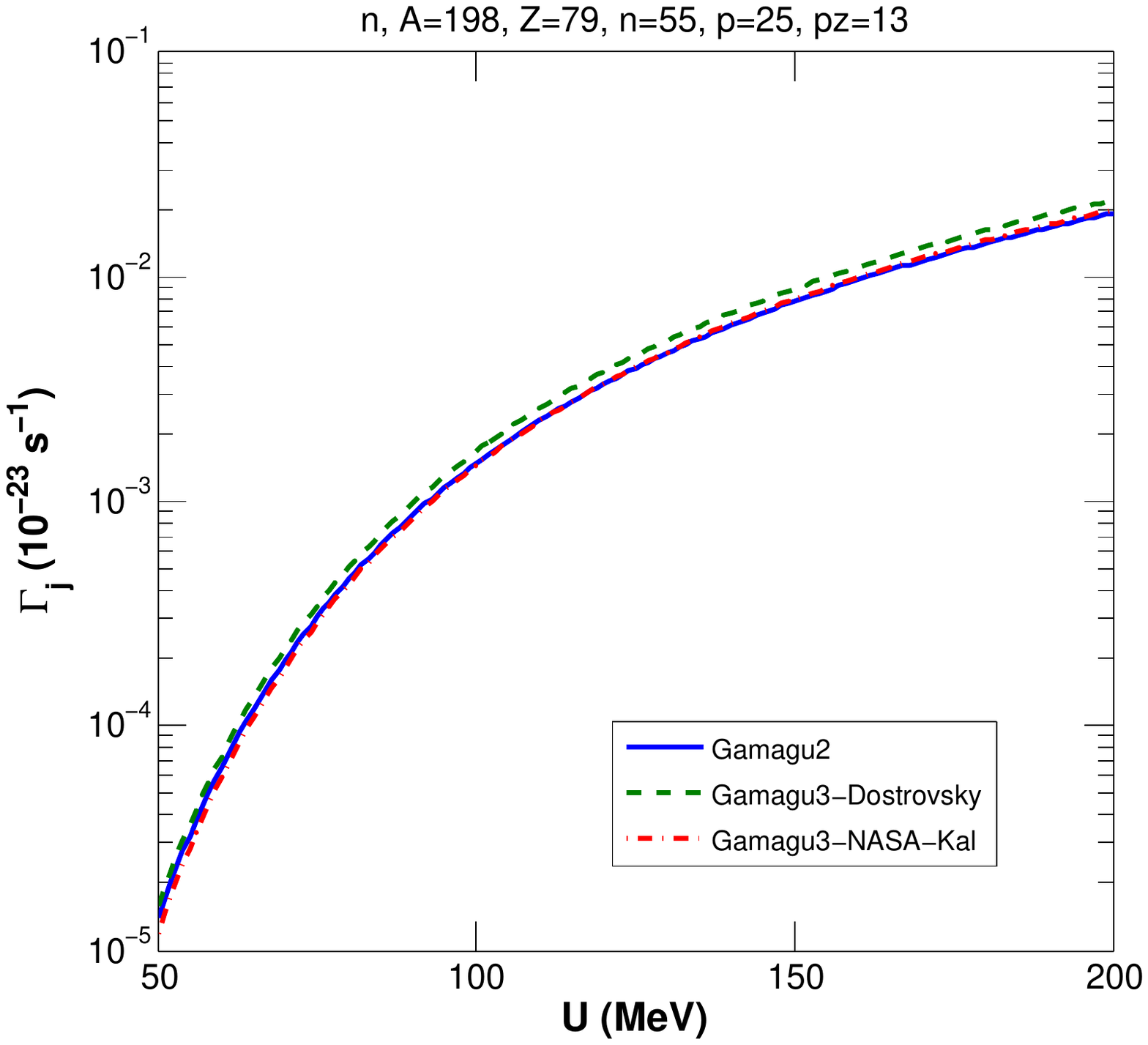}
\includegraphics[trim = 1.75in 2.in 1.75in 2.in, width=2.5in]{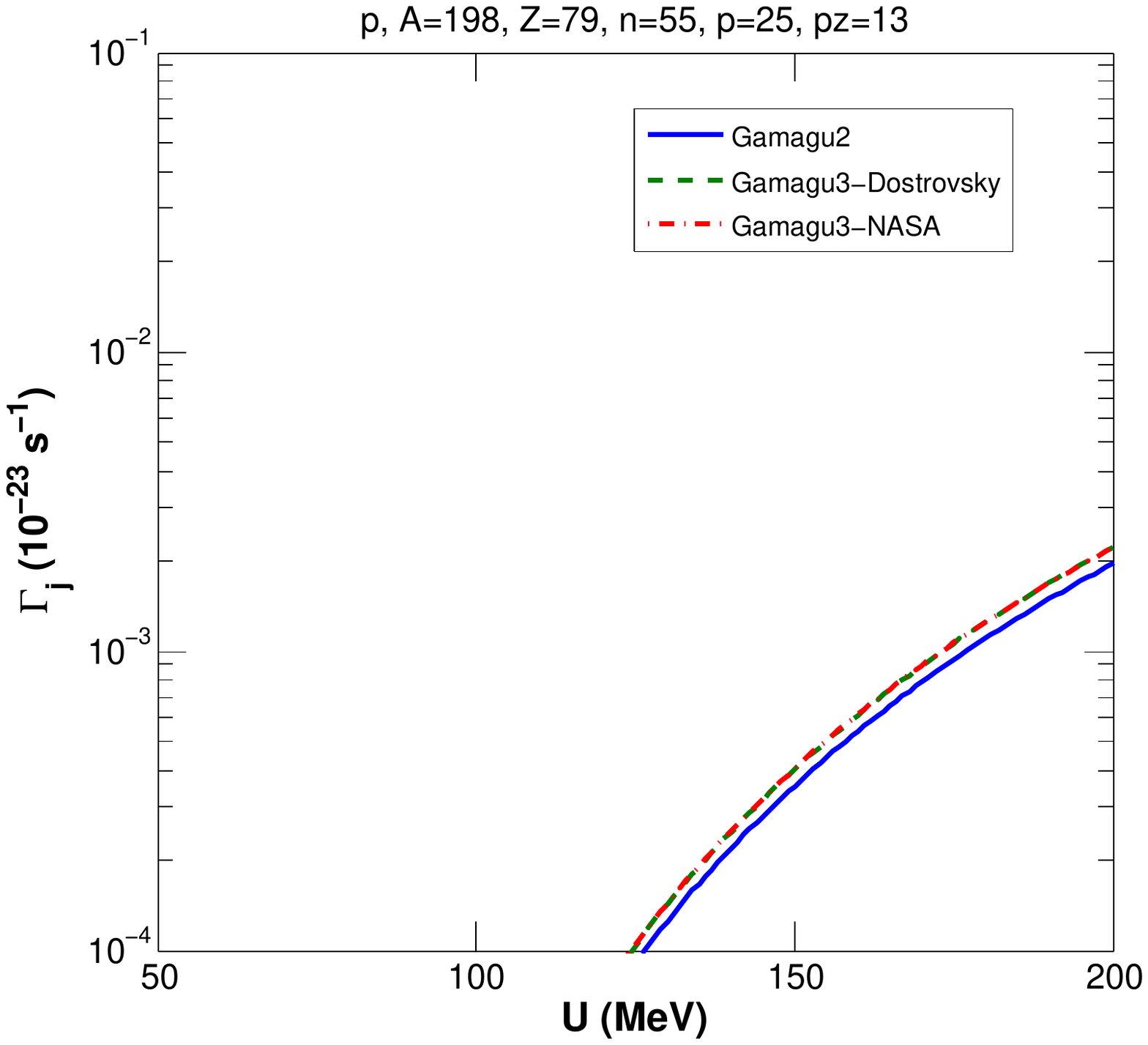}
\includegraphics[trim = 1.75in 2.5in 1.75in 3.in, width=2.5in]{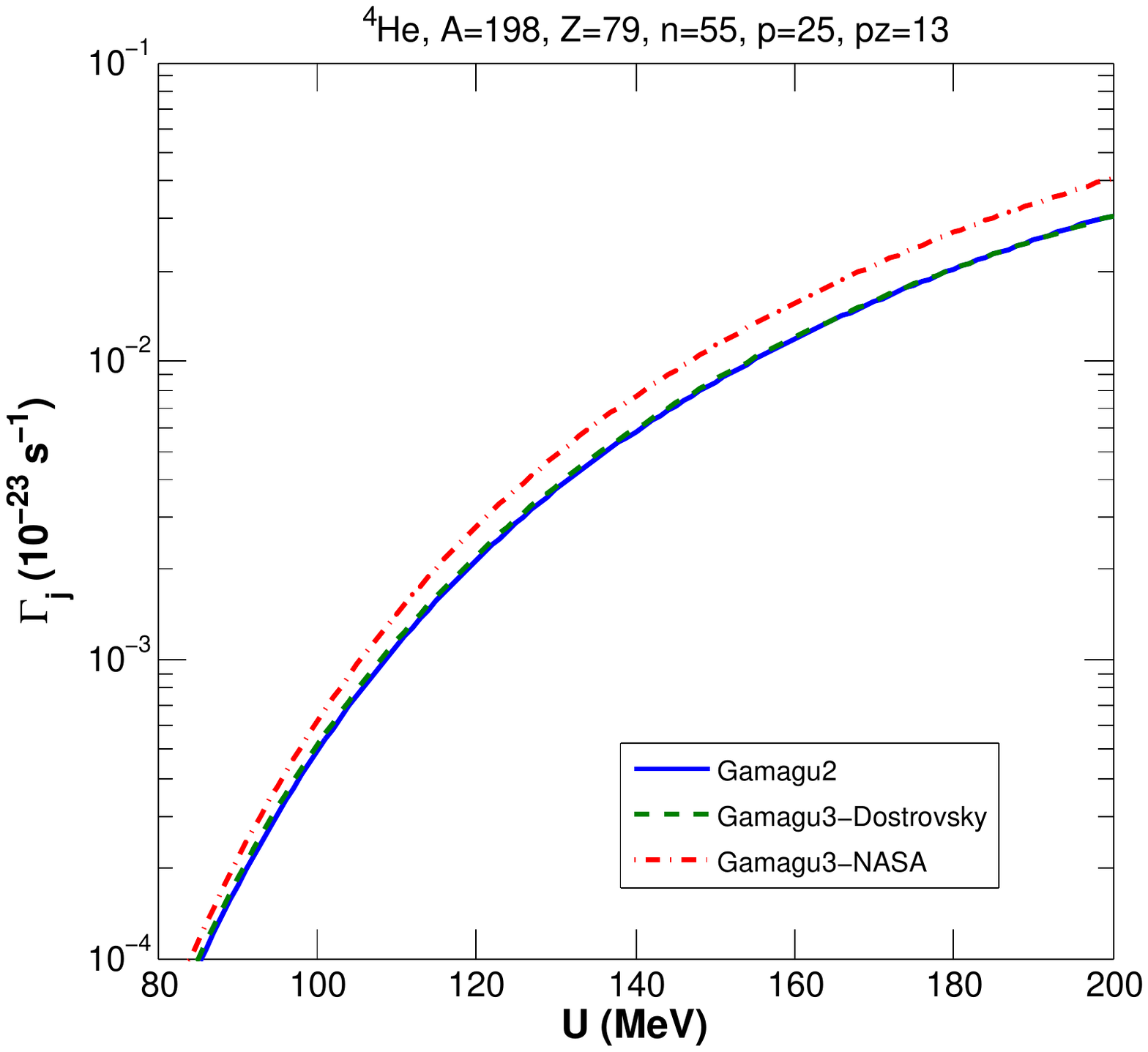}
\caption[]{$\Gamma_j$ as a function of the internal energy of the excited nucleus for emitted neutrons, protons, and $^4$He from an excited $^{198}$Au nucleus with 55 excitons, 25 particle excitons, and 13 charged particle excitons.}
\label{fig:Gamma_n_Kal}
\end{figure}

Fig.~\ref{fig:Gamma_n_Kal} shows the plots of $\Gamma_j$ as a function of the internal energy of the excited nucleus for emitted neutrons, protons, and $^4$He from an excited $^{198}$Au nucleus with 55 excitons, 25 particle excitons, and 13 charged particle excitons. ``Gamagu2'' shows the old CEM03.03 $\Gamma_j$ calculation results. ``Gamagu3'' shows the results of our new calculations, using either the Dostrovsky {\it et al.} or NASA inverse cross sections. Note that ``Gamagu2'' should be very similar to ``Gamagu3-Dostrovsky'' because the only significant difference is the method of integration. The proton and neutron $\Gamma_j$ differences between ``Gamagu2'' and ``Gamagu3-Dostrovsky'' arise from numerical integration used in our new FORTRAN routine {\sf gamagu3} of CEM03.03F versus an analytical calculation used in the CEM03.03 FORTRAN routine {\sf gamagu2}.

Better integration methods could be investigated at a later time. However, current integration methods are sufficient because individual $\Gamma_j$ precision is not extremely important for choosing what type of particle/LF $j$ will be emitted. In contrast to analytical preequilibrium models, the Monte Carlo method employed by our CEM uses the ratios of $\Gamma_j$ to the sum of $\Gamma_j$ over all $j$. That is, if we estimate all $\Gamma_j$ with the same percentage error, the final choice of the type $j$ of particle/LF to be emitted as simulated by CEM would be the same as if we would calculate all $\Gamma_j$ exactly. We think that this is the main reason why CEM provided quite reasonable results using the old Dostrovsky {\it et al.} approximation for inverse cross sections, in spite of the fact that, as we see from Figs.~\ref{fig:Mashnik}--\ref{fig:n+Pb_Kal}, individual inverse cross sections calculated with the Dostrovsky {\it et al.} method are not good enough in a large range of energies. The ratios $\Gamma_j/\sum_j(\Gamma_j$) were probably estimated well enough, providing a reasonable Monte Carlo sampling of $j$.

\subsection{Kinetic Energy Simulation}
Once a fragment type $j$ has been chosen for emission, the kinetic energy of this fragment needs to be determined. This is done by sampling the kinetic energy from the $\lambda_j$ distribution. Our new $\lambda_j$, with the NASA cross section, is:

\begin{equation}
\begin{split}
\lambda_j(p,h,E,T) = & \gamma_j \frac{2s_j + 1}{\pi^2\hbar^3} \mu_j \Re (p,h) \\
	& \times \frac{\omega (p-p_j,h,E-B_j-T)}{\omega (p,h,E)} \\
	& \times \frac{\omega (p_j,0,T+B_j)}{g_j} T \pi r^2_0 \\
	& \times (A_P^{1/3} + A_T^{1/3} + \delta_T )^2 (1 - R_c \frac{B_T}{T_{cm}})X_m (T) .
\end{split}
\label{eq:lambda_j_NASA}
\end{equation}

where $g_j$ is defined by Eq.~(\ref{eq:g_j}) and

\begin{equation}
\begin{split}
\delta_T =  1.85 S + \frac{0.16 S}{T_{cm}^{1/3}} - D[1 - e^{-T/T_1}] - 0.292 e^{-T/792} \\
	 + \frac{0.91(A_T - 2Z_T)Z_P}{A_T A_P} , \\
X_m = 1 - X_1 exp\left(\frac{-T} {X_1 (1.2 + 1.6[1 - exp {(T/15)}])}\right)  , \\
B_T = 1.44 Z_P Z_T / \left(r_P + r_T + { {1.2(A_P^{1/3} + A_T^{1/3})} \over {T_{cm}^{1/3}}} \right) .
\end{split}
\label{eq:NASA_details}
\end{equation}

The details of $r_0, R_c, S, D, T_1, r_P,$ and $r_T$ can be found in~\cite{NASAl}. Note that the NASA inverse cross sections contain dependences on both the lab-reference-frame kinetic energy ($T$) and the center-of-momentum-reference-frame kinetic energy ($T_{cm}$). The relativistic transformation between the two is not trivial. In addition, $T$ is in units of MeV/nucleon in the NASA inverse cross sections, while $T_{cm}$ is in units of MeV. The level density, $\omega$, also contains $T$-dependences, also in units of MeV. Finally, as noted above, for neutrons we use a NASA-Kalbach (``hybrid'') inverse cross section in place of the pure NASA approximation. To conclude, the energy-dependence of $\lambda_j$ for our new NASA-Kalbach inverse cross section approximation is very complicated, which affects the method we chose to sample $T_j$, as discussed below.

To sample $T_j$ uniformly from the $\lambda_j$ distribution using the Monte Carlo method, we must first find the maximum of $\lambda_j$. In CEM03.03, this is done analytically using the derivative of $\lambda_j$ with respect to $T_j$, due to the simple nature of the energy-dependence in the systematics by Dostrovsky {\it et al.}. As previously explained, however, the NASA cross section energy-dependence is extremely complicated and therefore we find the maximum of $\lambda_j$ numerically using the Golden Section method. This also provides us flexibility in the future to modify $\lambda_j$ without consequence to our kinetic energy module.

\begin{figure}[h!]
\centering
\includegraphics[trim = 2.25in 2.5in 2.25in 2.5in, width=2.5in]{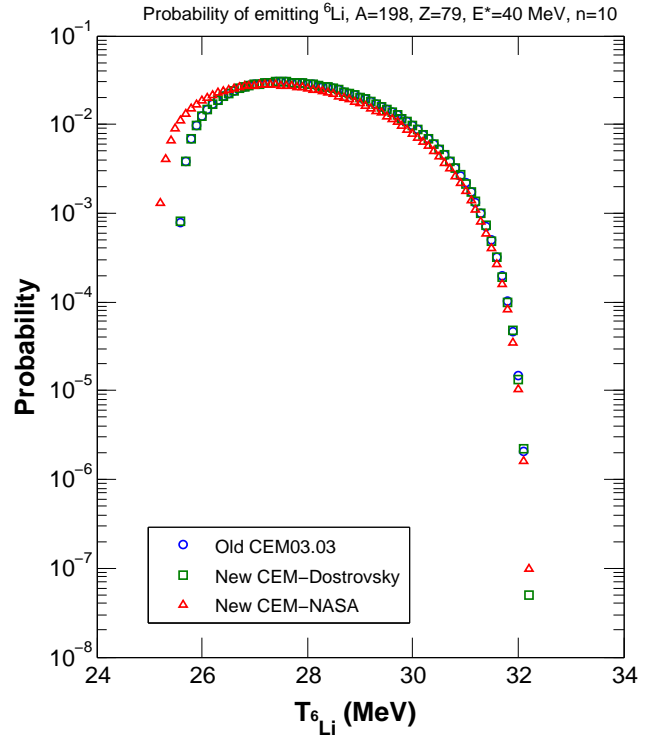}
\caption[]{The normalized probability of emitting $^6$Li with a given kinetic energy $T_{Li}$, simulated using the Monte Carlo method according to Eq.~\ref{eq:lambda_j_NASA} in the preequilibrium stage. The circles are results from the old kinetic energy subroutine; the squares are results from the new kinetic energy subroutine using the Dostrovsky {\it et al.} inverse cross section; the triangles are from the new kinetic energy subroutine using the NASA inverse cross section.}
\label{fig:kin_energy}
\end{figure}

After finding the maximum value of $\lambda_j$, the kinetic energy of the emitted fragment $j$ is uniformly sampled from the $\lambda_j$ distribution using a Gamma distribution (shape parameter $\alpha = 2$) as the comparison function. Fig.~\ref{fig:kin_energy} illustrates results for the probability of emitting $^6$Li with a given kinetic energy $T_{Li}$. Probabilities from the $\lambda_j$ distributions with the NASA inverse cross sections differ slightly from those with the Dostrovsky {\it et al.} inverse cross sections primarily because the NASA coulomb barriers are based on $T_{cm}$, as opposed to $T$.

\section{Results}
\begin{figure*}[ht!]
\centering
\includegraphics[width=7.0in]{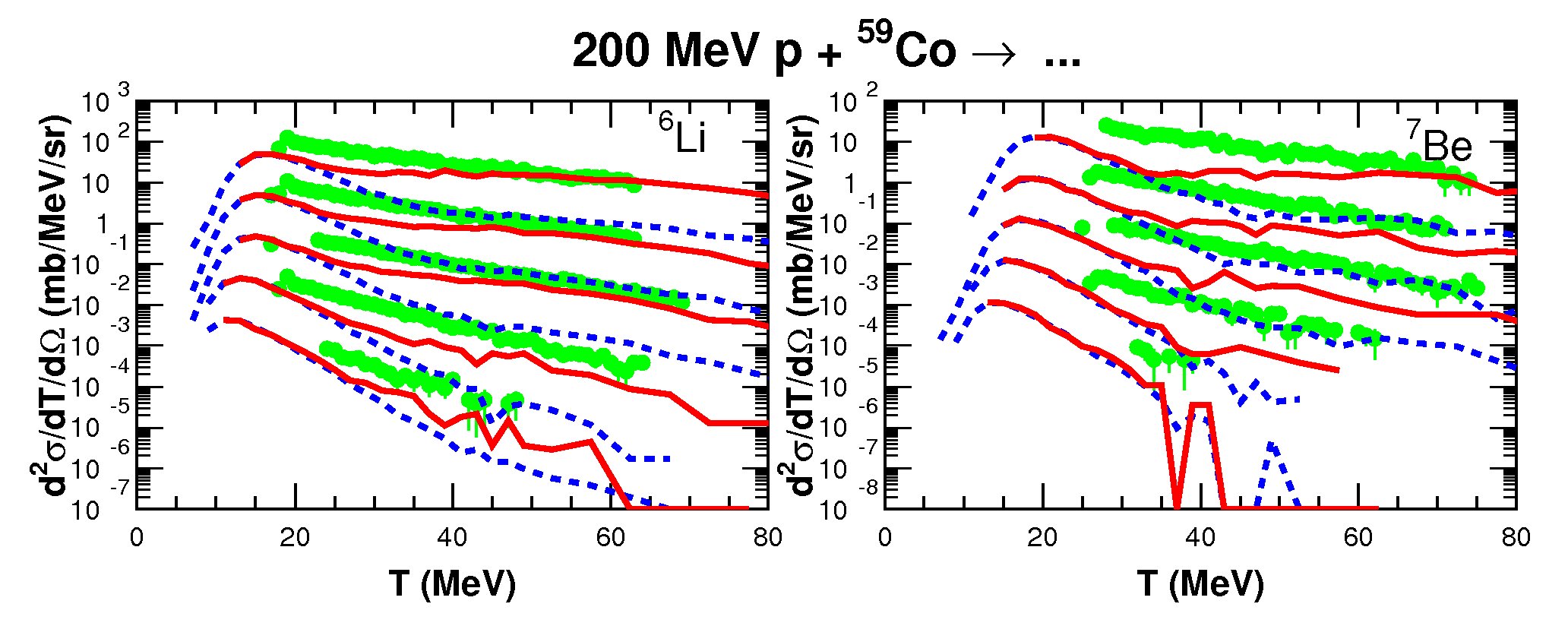}
\caption[]{Double differential cross section for the production of $^6$Li and $^7$Be from the reaction 200~MeV~p + $^{59}$Co, for the angles of 20\degree, 45\degree, 60\degree, 90\degree, and 110\degree. The 110\degree spectra (the lower sets) are shown unscaled, while the 90\degree, 60\degree, 45\degree, and 20\degree spectra are scaled up by successive factors of 10, respectively. The blue dashed lines are the expanded-MEM (i.e., CEM03.03F) results with the Dostrovsky {\it et al.} inverse cross sections, and the red solid lines are results by CEM03.03F with the upgraded NASA-Kalbach (i.e., ``hybrid'') inverse cross sections. The green circles are experimental data by Machner, {\it et al} \cite{Machner}.}
\label{fig:p200Co_new}
\end{figure*}

Our preliminary results are promising. Fig.~\ref{fig:p200Co_new} shows the double differential cross section for the production of $^6$Li and $^7$Be from the reaction 200~MeV~p + $^{59}$Co. Notice the improved agreement with data in the high-energy tails. This reaction also highlights the importance of eventually upgrading the inverse cross sections used in the evaporation stage of CEM as well. The evaporation stage produces the peak of the spectra, which for this reaction is too low, especially for $^7$Be. With the implementation of the NASA inverse cross sections in the preequilibrium stage we see improved agreement with data in the high-energy tails, but in order to achieve improved agreement in the peak we would need to also implement the NASA inverse cross sections in the evaporation stage. We plan to do this in the future.

\begin{figure*}[ht!]
\centering
\includegraphics[width=7.0in]{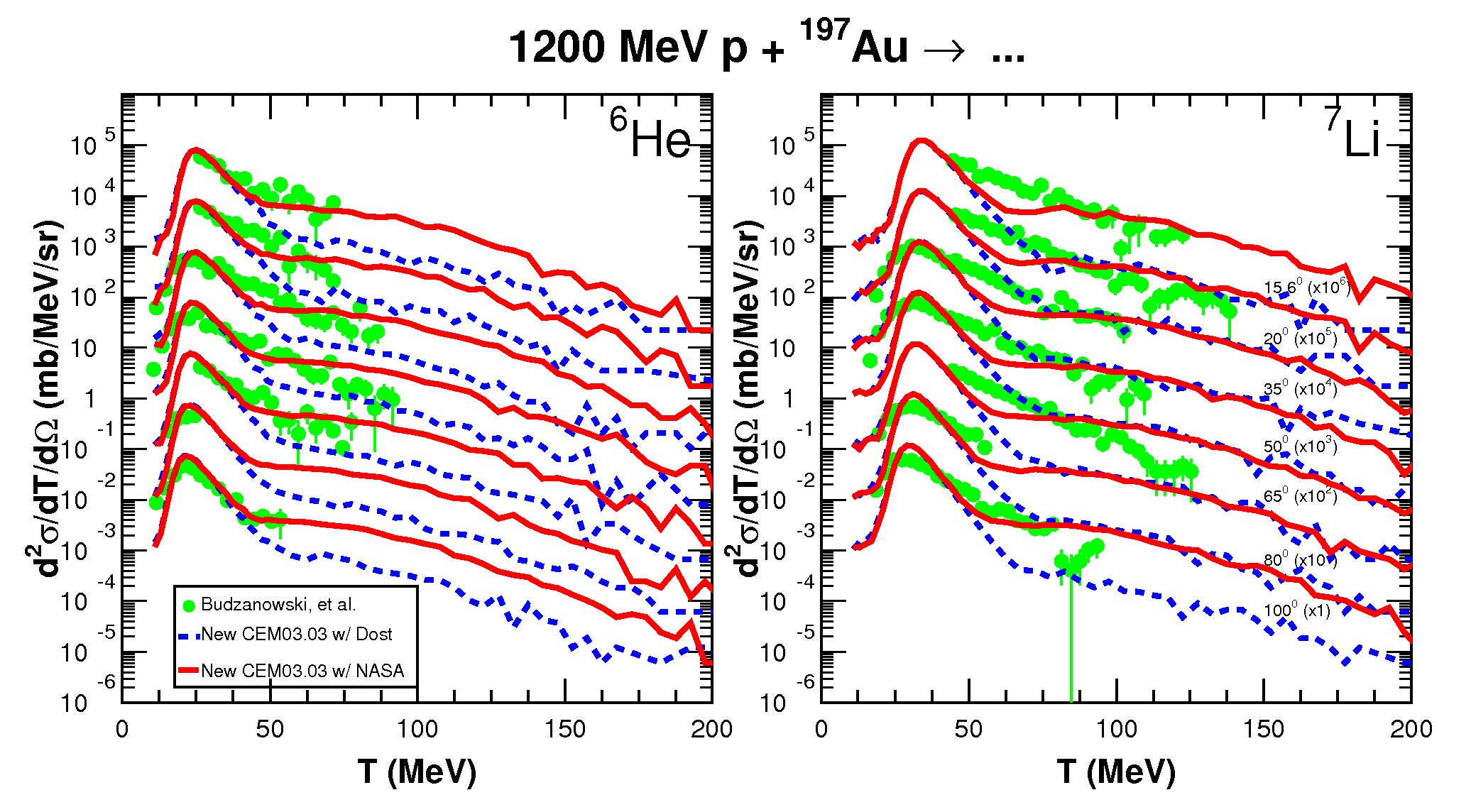}
\caption[]{Double differential cross section for the production of $^6$He and $^7$Li from the reaction 1200~MeV~p + $^{197}$Au, for the angles of 15.6\degree, 20\degree, 35\degree, 50\degree, 65\degree, 80\degree, and 100\degree. The 100\degree spectra (the lower sets) are shown unscaled, while the 80\degree, 65\degree, etc., down to 15.6\degree spectra are scaled up by successive factors of 10, respectively. The blue dashed lines are the expanded-MEM (i.e., CEM03.03F) results with the Dostrovsky {\it et al.} inverse cross sections, and the red solid lines are results by CEM03.03F with the upgraded NASA-Kalbach (i.e., ``hybrid'') inverse cross sections. The green circles are experimental data from Ref.~\cite{Budzanowski}.}
\label{fig:p1200Au_new}
\end{figure*}

\begin{figure}[h]
\centering
\includegraphics[width=3.5in]{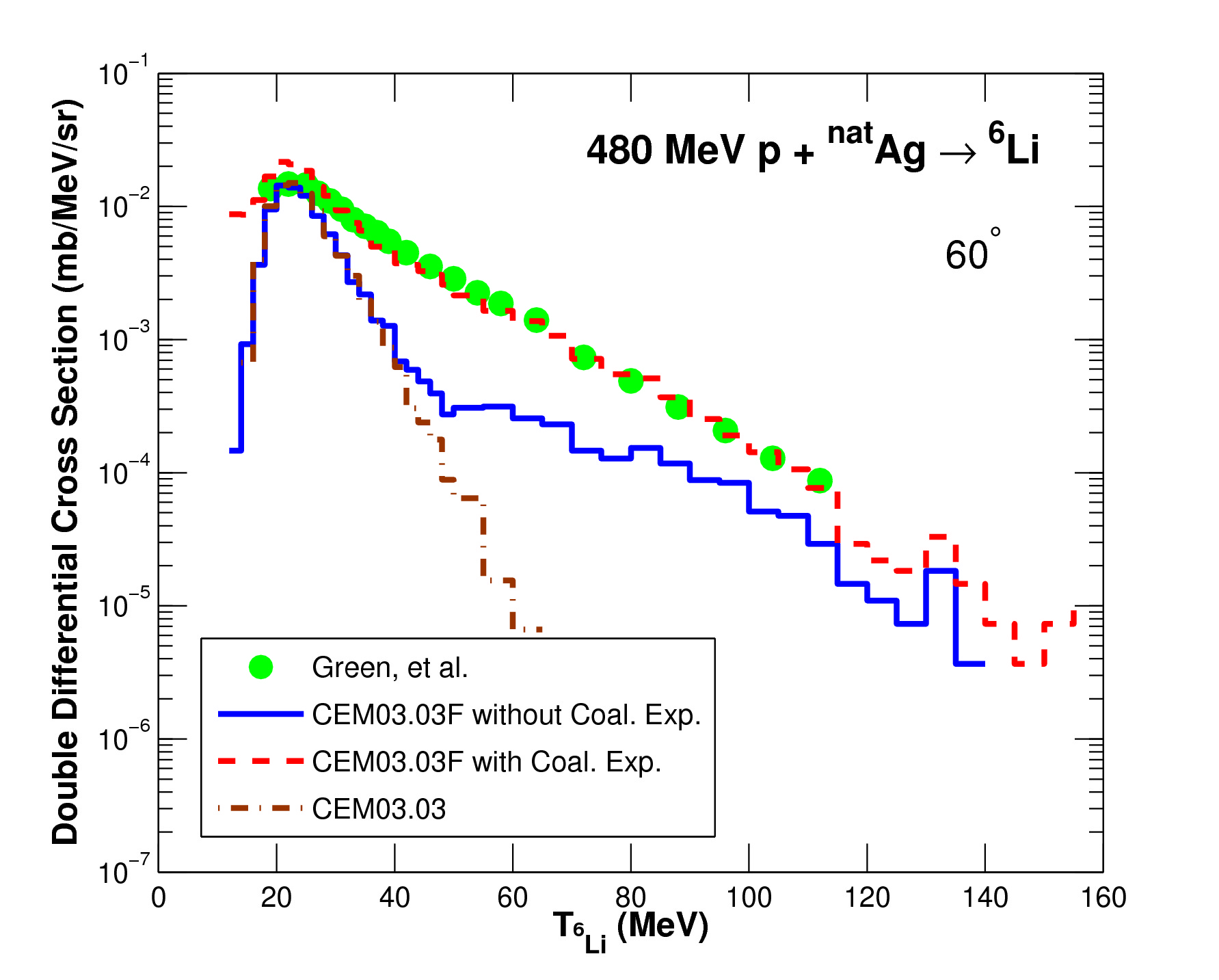}
\caption[]{Comparison of experimental results of the reaction 480 MeV p + $^{nat}$Ag $\rightarrow$ $^6$Li at 60\degree ~by Green {\it et al.} \cite{Green480} (green circles), with simulations from the original CEM03.03 (brown dashed-dotted lines), CEM03.03F without coalescence expansion (blue solid lines) and the CEM03.03F with coalescence expansion (red dashed lines).}
\label{fig:p480AgLi6}
\end{figure}

For another example of our results, Fig.~\ref{fig:p1200Au_new} displays the double differential cross section for the production of $^6$He and $^7$Li from the reaction 1200~MeV~p + $^{197}$Au. The blue dashed lines are the expanded-MEM (i.e., CEM03.03F) results with the Dostrovsky {\it et al.} inverse cross sections, and the red solid lines are results by CEM03.03F with the upgraded NASA-Kalbach (i.e., ``hybrid'') inverse cross sections. The green circles are experimental data from Ref.~\cite{Budzanowski}. We see an improved accuracy in the high-energy tails of spectra calculated with the NASA inverse cross sections, although some of the results are too hard and there is a dip in the spectra at ~50--75 MeV. We would like to note a recent paper by A. Boudard {\em et al.} \cite{Boudard2013}, which obtained very similar results for $^7$Li using INCL4.6 + ABLA07, and similar results for $^6$He but with a little lower evaporation peak. Current work is being undertaken to expand the coalescence model in CEM, which helps soften the spectra and smooth out the dip that appears around 50-75 MeV. This work is an ongoing process, but we display some of our preliminary results in Fig.~\ref{fig:p480AgLi6}. For further details, see Refs.~\cite{ANS2015, NUFRA2015}.

\section{Conclusion}
\label{sec-4}
The inverse cross section approximation in the preequilibrium and evaporation stages of CEM03.03 is based on the Dostrovsky {\it et al.} inverse cross section model. Better cross section systematics are available at present. We performed a comparison of several inverse cross section models and determined that the NASA (Tripathi, {\it et al.}) approximation is, in general, the most accurate when compared with experimental data. 

We implemented the NASA inverse cross section model into the preequilibrium stage of CEM03.03F. This included writing FORTRAN modules containing the NASA total reaction cross section and coulomb barrier approximations, adding Kalbach systematics for low-energy neutron inverse cross sections, re-writing the $\Gamma_j$ routines (including transforming them into modular FORTRAN), adding Gauss-Laguerre quadrature for cases of high exciton number, and re-modeling the selection of particle or light fragment kinetic energy. These technical improvements lead to greater flexibility and robustness, and future upgrades can be made easily. 

Our preliminary results are promising and indicate improved agreement with experimental data using the NASA inverse cross section model versus the Dostrovsky {\it et al.} approximation.

There are several implications of this work on MCNP6. CEM03.03 is the default event-generator in MCNP6 for high-energy collisions induced by nucleons, pions, and gammas at energies up to several GeVs. Improvements to the CEM inverse cross sections should, therefore, result in improved predictions of particle spectra and total production cross sections, especially above $\sim$100 MeV and for fragments heavier than $^4$He, among other results. 

MCNP6 uses the updated Barashenkov and Polanski total reaction cross section systematics to simulate the mean-free path of neutrons, protons, and light fragments up to $^4$He. It uses a parameterization based on a geometric cross section for fragments heavier than $^4$He. Possible direct improvement of MCNP6 may be obtained by replacing the Barashenkov and Polanski model with NASA systematics and by replacing the geometric cross section approach with the better NASA model. We hope to do this in the future.

Future recommendations include investigating adaptive quadrature and upgrading the inverse cross section model used in the evaporation stage to the NASA-Kalbach (hybrid) cross sections.

\vspace*{5mm}
{\noindent
{\bf Acknowledgments}
}

We are grateful to our colleagues, Drs. Konstantin K. Gudima and Arnold J. Sierk for a long and very fruitful collaboration with us and for several useful discussions of the results presented here.

We are grateful to Drs. Lawrence J. Cox and Avneet Sood of Los Alamos National Laboratory and to Prof. Akira Tokuhiro of the University of Idaho for encouraging discussions and support.

This study was carried out under the auspices of the National Nuclear Security Administration of the U.S. Department of Energy at Los Alamos National Laboratory under Contract No. DE-AC52-06NA25396.

This work is supported in part (for L.M.K) by the M. Hildred Blewett Fellowship of the American Physical Society, 
{\noindent
www.aps.org.
}

\vspace*{5mm} 
{\noindent
{\bf References}
}

\end{document}